\documentclass[preprint,aps,12pt,showpacs,nofootinbib,tightenlines]{revtex4}
\usepackage{amsmath}
\usepackage{amssymb}
\usepackage{epsfig}
\usepackage{graphicx}
\begin{document}

\newcommand{\beq}{\begin{eqnarray}}
\newcommand{\eeq}{\end{eqnarray}}
\newcommand{\non}{\nonumber\\ }

\newcommand{\acp}{{\cal A}_{CP}}
\newcommand{\etap}{\eta^{(\prime)} }
\newcommand{\etapr}{\eta^\prime }
\newcommand{\jpsi}{ J/\Psi }
\newcommand{\kst}{ K_0^*(1430) }
\newcommand{\kstb}{ \bar{K}_0^*(1430)^0 }
\newcommand{\kt}{ K_0^* }
\newcommand{\ktb}{ \bar{K_0^*} }
\newcommand{\kb}{ \bar{K}}

\newcommand{\pb}{\phi_{B}}
\newcommand{\pp}{\phi_{\pi}}
\newcommand{\pk}{\phi_{K}^A}
\newcommand{\pepr}{\phi_{\eta'}^A}
\newcommand{\pkp}{\phi_{K}^P}
\newcommand{\pep}{\phi_{\eta}^P}
\newcommand{\peprp}{\phi_{\eta'}^P}
\newcommand{\pkt}{\phi_{K}^T}
\newcommand{\pet}{\phi_{\eta}^T}
\newcommand{\peprt}{\phi_{\eta'}^T}
\newcommand{\fb}{f_{B} }
\newcommand{\fpi}{f_{\pi} }
\newcommand{\feta}{f_{\eta} }
\newcommand{\fetap}{f_{\eta'} }
\newcommand{\rpi}{r_{\pi} }
\newcommand{\re}{r_{\eta} }
\newcommand{\rep}{r_{\eta'} }
\newcommand{\mb}{m_{B} }
\newcommand{\mop}{m_{0\pi} }
\newcommand{\moe}{m_{0\eta} }
\newcommand{\moep}{m_{0\eta'} }

\newcommand{\psl}{ P \hspace{-1.8truemm}/ }
\newcommand{\nsl}{ n \hspace{-2.2truemm}/ }
\newcommand{\vsl}{ v \hspace{-2.2truemm}/ }
\newcommand{\epsl}{\epsilon \hspace{-1.8truemm}/\,  }

\def \epjc{ Eur. Phys. J. C }
\def \jpg{  J. Phys. G }
\def \npb{  Nucl. Phys. B }
\def \plb{  Phys. Lett. B }
\def \pr{  Phys. Rep. }
\def \prd{  Phys. Rev. D }
\def \prl{  Phys. Rev. Lett.  }
\def \zpc{  Z. Phys. C  }
\def \jhep{ J. High Energy Phys.  }

\title{$B \to  \kst K$ decays in the perturbative QCD approach
\footnote{This work is partially supported by the National Natural Science
Foundation of China under Grant No.10575052, 10605012 and
10735080.}}
\author{ Xin Liu\footnote{ hsinlau@126.com} and
Zhen-jun Xiao\footnote{ xiaozhenjun@njnu.edu.cn}
} \affiliation{ Department of Physics and Institute of Theoretical
Physics, Nanjing Normal University, Nanjing, Jiangsu 210097, P.R.
China}
\date{\today}
\begin{abstract}
In this article, we calculate the branching ratios of $B \to \kst K$
decays by employing the perturbative QCD (pQCD) approach at leading
order. We perform the evaluations in the two scenarios for the
scalar meson spectrum. We find that (a) the leading order pQCD
predictions for the branching ratio $Br(B^+ \to K^+
\bar{\kt}(1430)^0)$ which is in good agreement with the experimental
upper limit in both scenarios, while the pQCD predictions for other
considered  $B \to \kst K$ decay modes are also presented and will
be tested by the LHC experiments; (b) the annihilation contributions
play an important role in these considered decays, for $B^0 \to
{\kst}^\pm {K}^\mp$ decays, for example, which are found to be
$(1-4) \times 10^{-6}$.

\end{abstract}

\pacs{13.25.Hw, 12.38.Bx, 14.40.Nd}

\maketitle


It is well-known that the scalar meson spectrum is one of the
interesting topics for both experimental and theoretical studies,
but the underlying structure of the light scalar mesons is still
controversial. Perhaps, the $B \to SP$ decays can give us the
opportunity to receive new understanding on the scalar meson.

On the theory side, up to now, some two body non-leptonic B meson
involving a scalar $\kst$(For the sake of simplicity, we will use
$\kt$ to denote $\kst$ in the following section) meson decays have
been studied by using various theoretical methods or approaches, for
example, in Ref.~\cite{CCY06,epjc50,CCY08,liu092}, where the authors
investigated the properties of $\kt$ by calculating the branching
ratios, CP-violating asymmetries and other physical quantities. In
this paper, based on the assumption of two-quark structure of scalar
$\kt$ meson, we will calculate the branching ratios of $B^+ \to
{\kt}^+ {\kb}^0, K^+ \bar{\kt}^0$ and $B^0/\bar{B}^0 \to {\kt}^0
{\kb}^0, K^0 \bar{\kt}^0, {\kt}^+ K^-, K^+ {\kt}^-$ decays directly
by employing the low energy effective Hamiltonian \cite{buras96} and
the pQCD factorization approach~\cite{lb80,cl97,luy01,li2003}.

On the experimental side, only one upper limit on $B^+ \to
\bar{\kt}^0 K^+$ is available now~\cite{pdg2008,hfag08} (upper
limits at $90\%$ C.L.):
 \beq
 Br(B^+ \to \bar{\kt}^0 K^+) &<& 2.2 \times 10^{-6}\;.\label{eq:exp1}
 \eeq
But this situation will be improved rapidly when the LHC experiment starts to run in 2009.

This paper is organized as follows. In Sec.~\ref{sec:1}, we
calculate analytically the related Feynman diagrams and find the
various decay amplitudes for the studied decay modes. In
Sec.~\ref{sec:n-d}, we show the numerical results for the branching
ratios of $B \to \kt K$ decays. A short summary and some discussions
are also included in this section.

\section{Perturbative Calculations of $B \to \kt K$ decays}\label{sec:1}

In the pQCD factorization approach, the decay amplitude of $B\to \kt K$
decays can be written conceptually as the convolution,
\beq
{\cal A}(B \to \kt K) &\sim &
\int\!\! d x_1 d x_2 d x_3 b_1 d b_1 b_2 d b_2 b_3 d b_3 \non
&& \times \mathrm{Tr} \left [ C(t) \Phi_{B}(x_1,b_1)
\Phi_{\kt}(x_2,b_2) \Phi_{K}(x_3, b_3) H(x_i, b_i, t) S_t(x_i)\,
e^{-S(t)} \right ], \label{eq:a2}
\eeq
where the term ``$\mathrm{Tr}$" denotes the trace over Dirac and color indices.
$C(t)$ is the Wilson coefficient. The function $H(x_i,b_i,t)$ is the
hard part and can be calculated perturbatively, while $b_i$ is the
conjugate space coordinate of $k_{iT}$, and $t$ is the largest
energy scale in hard function. The function $\Phi_M$ is the wave
function which describes hadronization of the quark and anti-quark
to the meson $M$. The threshold function $S_t(x_i)$ smears the
end-point singularities on $x_i$. The last term, $e^{-S(t)}$, is the
Sudakov form factor which suppresses the soft dynamics effectively.

\begin{figure}[tb]
\vspace{-0.5cm} \centerline{\epsfxsize=13cm \epsffile{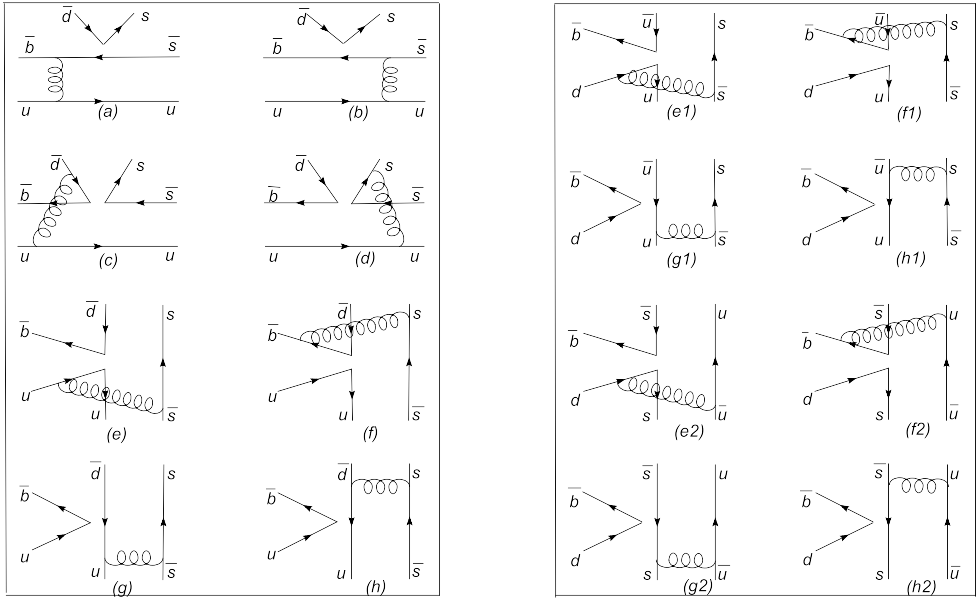}}
\vspace{-0.1cm} \caption{ Typical Feynman diagrams contributing to
 $B^+ \to {\kt}^+ {\kb}^0 (B^+ \to K^+ \bar{\kt}^0 )$(a-h in l.h.s.)
 and pure annihilation $B^0 \to {\kt}^+ K^-(B^0 \to K^+{\kt}^-)$(e1-h2 in r.h.s.)
 decays, respectively. }
 \label{fig:fig1}
\end{figure}

The low energy effective Hamiltonian for decay modes $B \to \kt K$
can be written as \beq \label{eq:heff} {\cal H}_{eff} = \frac{G_{F}}
{\sqrt{2}} \, \left[ V_{ub}^* V_{ud} \left (C_1(\mu) O_1(\mu) +
C_2(\mu) O_2(\mu) \right) - V_{tb}^* V_{td} \, \sum_{i=3}^{10}
C_{i}(\mu) \,O_i(\mu) \right] \; , \eeq where the Fermi constant
$G_{F}=1.166 39\times 10^{-5} GeV^{-2}$, $V_{ij}$ is the
Cabbibo-Kobayashi-Maskawa (CKM) matrix elements, $C_i(\mu)$ are the
Wilson coefficients at the renormalization scale $\mu$ and $O_i$ are
the four-fermion operators for the case of $\bar b \to \bar d $
transition.

The B meson is treated as a heavy-light system. We here use the same
B meson wave function as in Ref.~\cite{liu06,guodq07,xiao08a}, while
the treatment for the scalar meson $\kt$ is that same as in
Ref.~\cite{liu092}. For the distribution amplitudes of light
pseudoscalar $K$ meson, we directly adopt the form as given in
Ref.~\cite{ball}.

\begin{figure}[tb]
\vspace{0.1cm} \centerline{\epsfxsize=11 cm \epsffile{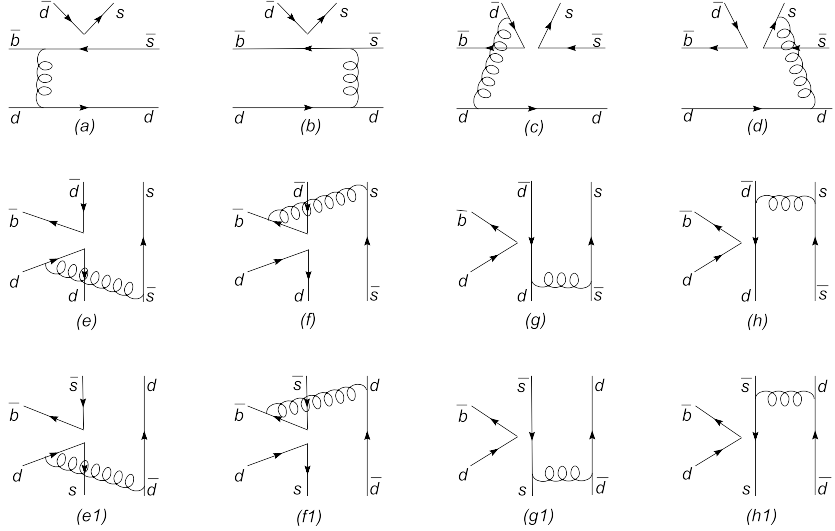}}
\vspace{-0.1cm} \caption{ Typical Feynman diagrams contributing to
$B^0 \to {\kt}^0 {\kb}^0 (B^0 \to \bar{\kt}^0 K^0)$ decays.}
\label{fig:fig2}
\end{figure}

At leading order, the relevant Feynman diagrams for the $B^+ \to
{\kt}^+ {\kb}^0, K^+ \bar{\kt}^0$, $B^0 \to {\kt}^+ K^-, K^+
{\kt}^-$ and $B^0 \to {\kt}^0 {\kb}^0, K^0 \bar{\kt}^0$ decays have
been shown in Figs.~\ref{fig:fig1} and \ref{fig:fig2}. Note that, on
the other hand, $\bar B^0$ meson can also decay into the same final
states ${\kt}^+ K^-, K^+ {\kt}^-$ and ${\kt}^0 {\kb}^0, K^0
\bar{\kt}^0$ simultaneously.

Based on the assumption of two quark structure of scalar $\kt$
meson, by analytical calculations of the relevant Feynman diagrams
and combining the contributions from different diagrams, one can
find the total decay amplitudes for the considered decays:
\beq {\cal
M}(B^+ \to {\kt}^+ \bar{K}^0)&=& -\xi_t \left\{f_K F_{e\kt}
(a_4-a_{10}/2) + f_K F_{e\kt}^{P2} (a_6-a_8/2)+M_{e\kt} (C_3-C_9/2)
\right. \non && \left. +M_{e\kt}^{P1} (C_5-C_7/2) + F^{P2}_{a\kt}
(a_6+a_8)+M_{a\kt}^{P1}(C_5+C_7)\right\} \non &&+ M_{a\kt}
\left\{\xi_u C_1- \xi_t (C_3+C_9)\right\}
 + F_{a\kt} \left\{\xi_u
a_1 - \xi_t (a_4+a_{10})\right\} \label{eq:ktpk0b}\\ {\cal M}(B^0
\to {\kt}^0 \bar{K}^0)&=&-\left\{ (a_4-a_{10}/2) \xi_t f_K F_{e\kt}
+(a_6-a_8/2) (\xi_t f_K F_{e\kt}^{P2}+\xi_t F_{a\kt}^{P2})
+(C_3\right.\non && \left.-C_9/2)\xi_t M_{e\kt}+(C_5-C_7/2) (\xi_t
 M_{e\kt}^{P1}+\xi_t
M_{a\kt}^{P1}) +\left(C_3+C_4\right.\right.\non
&&\left.\left.-(C_9+C_{10})/2\right)\xi_t M_{a\kt} +(C_6-C_8/2)
(\xi_t M_{a\kt}^{P2}+\xi_t M_{a\bar{K}}^{P2}) \right.\non &&
\left.+\left(a_3+a_4-a_5+(a_7-a_9-a_{10})/2\right)\xi_t F_{a\kt}
+(C_4-C_{10}/2)\xi_t \right.\non && \left.\cdot M_{a\bar{K}}+\xi_t
F_{a\bar{K}} \left(a_3-a_5+(a_7-a_9)/2\right)\right\}
\;,\label{eq:b0-b0b2kt0k0b}\\
{\cal M}(B^0 \to {\kt}^+ {K}^-)&=& M_{a{\kt}}\left\{\xi_u C_2 -
\xi_t (C_4+C_{10})\right\} -(C_6+C_8) \xi_tM_{a{\kt}}^{P2}-(C_4 \non
&&-C_{10}/2) \xi_t M_{a{K}}
 +F_{a{\kt}} \left\{\xi_u a_2 - \xi_t
(a_3-a_5-a_7+a_9)\right\}  \non &&-(C_6-C_8/2) \xi_tM_{a{K}}^{P2}-
\left(a_3-a_5+(a_7-a_9)/2\right)\xi_tF_{a{K}}
\label{eq:b0-b0b2ktpkm}
\eeq
where $\xi_u= V^*_{ub} V_{ud}$, $\xi_t= V^*_{tb} V_{td}$. The
individual decay amplitudes for $B \to \kt K$ decays, such as
$F_{e\kt}$ and $F_{e\kt}^{P2}$, etc, are similar to those
for $B \to \kt \etap$ decays as given in Ref.~\cite{liu092},
and can be obtained easily by the replacement of $\etap \rightarrow K$.

The Wilson coefficients $a_i$ in
Eq.~(\ref{eq:ktpk0b}-\ref{eq:b0-b0b2ktpkm}) are the combinations of
the ordinary Wilson coefficients $C_i(\mu)$,
\beq
a_1&=&C_2+C_1/3,\quad a_2=C_1+C_2/3,\quad a_i= C_i+C_{i\pm
1}/3,\quad i=3-10.\eeq
where the upper (lower) sign applies, when $i$ is odd (even).

The expressions of total decay amplitudes for $B^+ \to \bar{\kt}^0
K^+$ and $B^0 \to  \bar{\kt}^0 K^0, {K}^+ {\kt}^-$ modes can be
easily obtained with the replacement of $\kt\rightarrow K, \kb
\rightarrow \bar{\kt}$[here, ${\kt}(K)$ and $\bar{\kt}(\kb)$ denote
${\kt}^{+,0}(K^{+,0})$ and ${\kt}^-,\bar{\kt}^0(K^-,{\kb}^0)$] in
Eq.~(\ref{eq:ktpk0b},\ref{eq:b0-b0b2kt0k0b},\ref{eq:b0-b0b2ktpkm}),
respectively.

\section{Numerical results and Discussions}\label{sec:n-d}

For numerical calculation, we will use the following input
parameters: \beq
 \Lambda_{\overline{\mathrm{MS}}}^{(f=4)} &=& 0.250 {\rm GeV}, \quad
 f_K = 0.16 {\rm GeV}, \quad  f_{B} = 0.190 {\rm GeV},  \non
 m_0^{K}&=& 1.6 {\rm GeV},\quad m_{\kt}=1.425 {\rm GeV}, \quad M_W = 80.41{\rm
 GeV}, \non
 M_{B} &=& 5.28 {\rm GeV}, \quad
  \tau_{B^{\pm}}=1.638\times10^{-12}{\rm s},  \quad \tau_{B^{0}}=1.53\times10^{-12}{\rm s}.
 \label{para}
\eeq

For the CKM matrix elements, here we adopt the Wolfenstein
parametrization for the CKM matrix, and take $\lambda=0.2257,
A=0.814, \bar{\rho}=0.135$ and $\bar{\eta}=0.349$  \cite{pdg2008}.

In the two-quark picture of the scalar meson $\kt$, there are two
scenarios for the choice of the decay constants $f_{\kt}$, $\bar
f_{\kt}$ and the Gegenbauer moments $B_1$ and $B_3$ \cite{CCY06}:
\beq f_{\kt}&=&-0.025 \pm 0.002 {\rm GeV},\quad
\bar{f}_{\kt}=-0.300\pm 0.030 {\rm Gev}, \non B_1&=&0.58\pm0.07,\ \
\ \ \ \ \ \ \ \ \ \ \ \quad B_3=-1.20\pm0.08, \eeq in Scenario I,
and \beq f_{\kt}&=&0.037 \pm 0.004 {\rm GeV}, \quad
\bar{f}_{\kt}=0.445\pm 0.050{\rm  Gev},\non B_1&=&-0.57\pm0.13,\ \ \
\ \ \ \ \  \quad B_3=-0.42\pm0.22, \eeq in Scenario II~\cite{CCY06}.
In the numerical calculations we will consider these two scenarios,
respectively.

\begin{table}[htb]
\caption{The leading order pQCD predictions for the branching ratios
(in unit of $10^{-6}$) of $B \to \kt K$ decays in both scenarios,
where the numbers in parentheses are the central values of branching
ratios without the inclusion of annihilation diagrams. For
comparison, we also cite the experimental upper limit as given in
Ref.~\cite{pdg2008,hfag08}.} \label{tab:brI}
\begin{center}\vspace{-0.5cm}
\begin{tabular}[t]{c|c|c|c} \hline  \hline
Modes  & Scenario I   & Scenario II  & Data
\\ \hline
 $B^+ \to {\kt}^+ \bar{K}^0$     & $1.5^{+0.7+0.3+0.1+0.4}_{-0.4-0.2-0.1-0.2}(2.3) $ & $5.0^{+1.8+0.4+0.9+1.2}_{-1.2-0.3-0.6-1.0}(5.1) $ & $-$ \\
 $B^+ \to \bar{\kt}^0 K^+$            & $1.2^{+0.2+0.1+0.1+0.2}_{-0.1-0.1-0.1-0.2}(0.8) $ & $2.2^{+0.6+0.2+0.4+0.5}_{-0.4-0.2-0.1-0.4}(1.8)$ & $<2.2$ \\
 $B^0/\bar{B}^0 \to {\kt}^0 \bar{K}^0$& $2.7^{+0.5+0.5+0.4+0.6}_{-0.3-0.4-0.4-0.5}(2.7) $ & $7.5^{+2.1+0.5+1.7+1.7}_{-1.5-0.6-1.2-1.7}(6.0)$ & $-$ \\
 $B^0/\bar{B}^0 \to \bar{\kt}^0 K^0$  & $2.8^{+0.2+0.4+0.5+0.6}_{-0.1-0.4-0.4-0.5}(1.2) $ & $5.0^{+0.8+0.7+1.8+1.2}_{-0.6-0.6-1.0-1.0}(2.7)$ & $-$ \\
 $B^0 \to {\kt}^0 \bar{K}^0+\bar{\kt}^0 K^0$  & $5.1^{+1.0+0.8+0.8+1.1}_{-0.6-0.7-0.8-1.0}(5.3) $ & $14.9^{+4.3+1.3+2.7+3.5}_{-2.9-1.3-1.8-3.2}(12.0) $ & $-$ \\
 $B^0/\bar{B}^0 \to {\kt}^- K^+$ & $3.7^{+0.4+0.5+0.5+0.8}_{-0.3-0.4-0.4-0.7}(0.0) $ & $1.8^{+0.3+0.4+1.5+0.5}_{-0.2-0.3-0.9-0.4}(0.0)$ & $-$ \\
 $B^0/\bar{B}^0 \to {\kt}^+ K^-$ & $1.1^{+0.2+0.5+0.1+0.2}_{-0.2-0.4-0.2-0.2}(0.0) $ & $1.6^{+0.4+0.4+0.8+0.4}_{-0.2-0.5-0.5-0.3}(0.0) $ & $-$ \\
 $B^0 \to {\kt}^+ K^- +{\kt}^- K^+$ & $2.4^{+0.2+0.1+0.4+0.6}_{-0.1-0.1-0.4-0.4}(0.0) $ & $1.2^{+0.2+0.1+1.2+0.3}_{-0.1-0.1-0.6-0.2}(0.0) $ & $-$ \\ \hline \hline
\end{tabular}
\end{center}
\end{table}

Using the decay amplitudes obtained in last section, it is
straightforward to calculate the branching ratios for $B \to \kt K$
decays. From the leading order pQCD predictions for these considered
decays as displayed in Table~\ref{tab:brI}, some phenomenological
discussions are in order:

\begin{itemize}

\item[]{(1)}
It is worth stressing that the theoretical predictions in the pQCD
approach have relatively large theoretical errors induced by the
still large uncertainties of many input parameters. As shown in
Table~\ref{tab:brI}, in our pQCD predictions, the first error arises
from the B meson wave function shape parameter $\omega_b=0.40 \pm
0.04$. The second error is induced by the combination of the
uncertainties of Gegenbauer moments $a_1^K=0.17\pm 0.17$ and/or
$a_2^{K}=0.115\pm 0.115$. The last two errors come from the
combinations of the Gegenbauer coefficients $B_1$ and/or $B_3$, and the
decay constants $f_{\kt}$ and/or $\bar{f}_{\kt}$ of the scalar meson
$\kt$, respectively.

\item[]{(2)}
For $B^+ \to K^+ \bar{\kt}^0$ mode, one can find the the pQCD
prediction for the CP-averaged branching ratio agrees with the
currently available experimental upper limit in both scenarios.

\item[]{(3)}
For the charged $B^+ \to {\kt}^+ {\kb}^0$ and $B^+ \to K^+
\bar{\kt}^0$ channels, the CP-averaged branching ratios show us the
different features in two scenarios: the values are approximately
equal to each other for these two decays in Scenario I, while the
former is twice larger than the latter in Scenario II. We also show
the central values of the branching ratios with neglecting the
annihilation contributions as given in Table~\ref{tab:brI}, one can
see the difference between these considered two modes: the
annihilated diagrams are destructive to $B^+ \to {\kt}^+ {\kb}^0$
but constructive to $B^+ \to K^+ \bar{\kt}^0$ decays. Additionally,
the annihilation contributions play a more important role in
scenario I than that in Scenario II.

\item[]{(4)}
It is a little complicate for us to calculate the branch ratios of
$B^0/\bar B^0 \to f(={\kt}^0 {\kb}^0,{\kt}^+ K^-)(\bar f[=K^0
\bar{\kt}^0,K^+ {\kt}^-])$, since both $B^0$ and $\bar B^0$ can
decay into the same final state $f$ and $\bar f$ simultaneously.
Because of $B^0-\bar B^0$ mixing, it is very difficult to
distinguish $B^0$ from $\bar B^0$. But it is easy to identify the
final states. We therefore sum up $B^0/\bar B^0 \to {\kt}^0 {\kb}^0$
as one channel, and $B^0/\bar B^0 \to K^0 \bar{\kt}^0$ as another,
although the summed up channels are not charge conjugate
states~\cite{ly02}. Similarly, we have $B^0/\bar B^0 \to {\kt}^+
K^-$ as one channel, and $B^0/\bar B^0 \to K^+ {\kt}^-$ as another.
We also define the average branching ratio of the two channels
following the same convention as experimental
measure~\cite{pdg2008,hfag08}: $B^0 \to {\kt}^0 {\kb}^0 +K^0
\bar{\kt}^0$ and $B^0 \to {\kt}^+ {K}^- +K^+ {\kt}^-$. The branching
ratios for $B^0/\bar B^0 \to {\kt}^0 {\kb}^0, K^0 \bar{\kt}^0$,
$B^0/\bar B^0 \to {\kt}^+ K^-, K^+ {\kt}^-$, $B^0 \to {\kt}^0
{\kb}^0 +K^0 \bar{\kt}^0$ and $B^0 \to {\kt}^+ {K}^- +K^+ {\kt}^-$
decays have already been presented in Table~\ref{tab:brI}.

\item[]{(5)}
The branching ratios for $B^0/\bar B^0 \to {\kt}^0 {\kb}^0 (K^0
\bar{\kt}^0)$ in Scenario II are larger than those in Scenario I. As
for the annihilation corrections, one can see that they are
constructive to $B^0/\bar B^0 \to {\kt}^0 {\kb}^0$ nearly 0-20\%
and $B^0/\bar B^0 \to {K}^0 \bar{\kt}^0$ around 50\%, respectively.
By comparison, we find that the annihilation amplitudes are
important in both scenarios for $B^0/\bar B^0 \to {K}^0 \bar{\kt}^0$
decay while more important in Scenario II than that in Scenario I
for $B^0/\bar B^0 \to {\kt}^0 {\kb}^0$ decay. For the branching
ratio of $B^0 \to {\kt}^0 {\kb}^0 +K^0 \bar{\kt}^0$, we find that
the value in Scenario II is nearly three times as large as that in
Scenario I, however, the annihilation contributions are destructive
to the branching ratio in Scenario I while constructive to it in
Scenario II and play a more important role in Scenario II than that
in Scenario I.

\item[]{(6)}
From the pQCD predictions for the pure annihilation contributions
$B^0/\bar B^0 \to {\kt}^+ {K}^- (K^+ {\kt}^-)$ and $B^0 \to {\kt}^+
{K}^- +K^+ {\kt}^-$ as shown in last three lines of
Table~\ref{tab:brI}, we find that the leading order pQCD branching
ratios from this part can amount to $(1-4)\times 10^{-6}$, which
indicate the large annihilation effects in $B\to \kt K $ decays in
contrast to $B \to KK$~\cite{CL00} and $B \to K K^*$~\cite{GXX07}
decays. The branching ratio in Scenario I is about twice as large as
that in Scenario II for $B^0/\bar B^0 \to {\kt}^+ K^-$ while smaller
than that in Scenario II for $B^0/\bar B^0 \to K^+ {\kt}^-$. As for
the average of the two, the numerical prediction for $B^0 \to
{\kt}^+ {K}^- +K^+ {\kt}^-$ in Scenario II is half of that in
Scenario I.

\item[]{(7)}
Except for $B^+ \to K^+ \bar{\kt}^0$ decay, where an upper limit
is available now, there are no any experimental measurements for other decays
considered here.
We  therefore do not know which scenario is better now.
The pQCD predictions for the branching
ratios of $B \to \kt K$ decays will be tested by the LHC experiments.

\end{itemize}


In short, based on the assumption of two quark structure of scalar
$\kt$ meson, we calculated the branching ratios of $B \to \kt K$
decays at the leading order by using the pQCD factorization
approach.
From numerical calculations and
phenomenological analysis, we found that the pQCD predictions for
$Br(B^+ \to K^+ \bar{\kt}^0)$ is consistent with the existing
experimental upper limit in both scenarios.
We also predicted the
branching ratios for other decay channels. All of these predictions
will be tested by the LHC experiments. In the considered
$B \to \kt K$ decays, the annihilation contributions played an
important role, for $B^0 \to {\kt}^\pm
{K}^\mp$ modes, for example, which amount to $(1-4)\times 10^{-6}$.


\end{document}